\documentclass[%
 reprint,
 amsmath,amssymb,
 aps,
pra,
]{revtex4-1}

\usepackage{multirow}

\usepackage{graphicx}
\usepackage{dcolumn}
\usepackage{bm}

\usepackage{hyperref}
\hypersetup{
    colorlinks=true,
    linkcolor=blue, 
    citecolor=blue,
    urlcolor=blue
}

\usepackage{physics}
\usepackage{float}

\usepackage{numprint}

\graphicspath{{figures/}}

\usepackage{xifthen}

\usepackage{xcolor}

\usepackage{epstopdf}

\usepackage{multirow}
\usepackage[normalem]{ulem}
\definecolor{rulecolor}{RGB}{0,71,171}

\definecolor{tableheadcolor}{gray}{0.92}

\newcommand{\Xstate}{X^1\Sigma^+}
\newcommand{\Astate}{A^1\Pi}
\newcommand{\apistate}{a^3\Pi}

\newcommand{\XAtransition}{$\change{\Astate \leftarrow \Xstate}$}

\newcommand{\AXtransition}{$\Astate \rightarrow \Xstate$}

\newcommand{\Xstatevib}[1]{\Xstate (v\!=\!#1)}
\newcommand{\Astatevib}[1]{\Astate (v'\!=\!#1)}

\newcommand{\XAtransitionvib}[2]{$\change{\Astatevib{#1} \leftarrow \Xstatevib{#2}}$}

\newcommand{\Xapitransition}{$\change{\apistate \leftarrow \Xstate}$}

\newcommand{\alcl}[2][]{\ifthenelse{\equal{#1}{}}{Al$^{#2}$Cl}{$^{#1}$Al$^{#2}$Cl}}

\newcommand{\rfig}[1]{Fig.~\ref{#1}}
\newcommand{\rsec}[1]{Section~\ref{#1}}
\newcommand{\req}[1]{Eq.~(\ref{#1})}
\newcommand{\rtab}[1]{Tab.~\ref{#1}}

\newcommand{\kB}{k_\textrm{B}}

\newcommand{\change}[1]{{\color{black} #1}}
\newcommand{\remove}[1]{\ignorespaces}

\begin{document}


\title{Spectroscopy on the \XAtransition \hspace{1mm}Transition of Buffer-Gas Cooled AlCl}

\author{John R.~Daniel}%
\email{jdani017@ucr.edu}
\affiliation{Department of Physics and Astronomy, University of California, Riverside, CA 92521, USA}

\author{Chen Wang}%
\affiliation{Department of Physics and Astronomy, University of California, Riverside, CA 92521, USA}

\author{Kayla Rodriguez}%
\affiliation{Department of Physics and Astronomy, University of California, Riverside, CA 92521, USA}

\author{Taylor N.~Lewis}%
\affiliation{Department of Chemistry, University of California, Riverside, CA 92521, USA}

\author{Alexander Teplukhin}%
\affiliation{Theoretical Division (T-1, MS B221), Los Alamos National Laboratory, Los Alamos, New Mexico 87545, USA}

\author{Brian K.~Kendrick}%
\affiliation{Theoretical Division (T-1, MS B221), Los Alamos National Laboratory, Los Alamos, New Mexico 87545, USA}

\author{Christopher Bardeen}%
\affiliation{Department of Chemistry, University of California, Riverside, CA 92521, USA}

\author{Boerge Hemmerling}%
\affiliation{Department of Physics and Astronomy, University of California, Riverside, CA 92521, USA}

\date{\today}

\begin{abstract}
Aluminum monochloride (AlCl) has been proposed as an excellent candidate for laser cooling.
Here we present absorption spectroscopy measurements on the \XAtransition\ transition in AlCl inside a cryogenic helium buffer-gas beam cell. The high resolution absorption data enables a rigorous, quantitative comparison with our high-level {\it ab initio} calculations of the electronic and rovibronic energies, providing a comprehensive picture of the AlCl quantum structure. The combination of high resolution spectral data and theory permits the evaluation of spectroscopic constants and associated properties, like equilibrium bond length, with an order of magnitude higher precision. Based on the measured molecular equilibrium constants of the $\Astate$ state, we estimate a Franck-Condon factor of the \XAtransition\ of 99.88\%, which confirms that AlCl is amenable to laser cooling. 
\end{abstract}

\maketitle

\section{Introduction}
\label{sec:intro}

Research into cold and ultracold molecules is driven by their potential to enable a variety of novel applications \cite{McCarron2018a,Fitch2021}, including 
precision measurements and searches for new physics \cite{ACMECollaboration2014,Fitch2021a,Cairncross2017,Kozyryev2018,Yu2021,Hutzler2020,ORourke2019,Aggarwal2018},
controlled chemical reactions \cite{Krems2008,Ni2010,Ye2018,Ospelkaus2010},
quantum computing \cite{DeMille2002,Yelin2006,Yu2019}, and
quantum simulations \cite{Carr2009,Micheli2006}.
While the rich internal structure of molecules is at the heart of these applications, the complexity of the molecular structure renders traditional atom laser cooling and trapping techniques challenging.
Experiments which produce ultracold molecules by associating laser cooled ultracold atoms have seen enormous progress over the last decade \cite{Sage2005,Ni2008,Danzl2010,Aikawa2010,Takekoshi2014,Molony2014,Park2015,Guo2016,Liu2018,DeMarco2019}. To access a large set of chemically diverse molecules, e.g.~fluorides or chlorides whose constituents are not amenable to laser cooling, direct laser cooling can be employed to a certain set of molecules.
At present, several diatomic species have been laser cooled directly and confined in a magneto-optical trap
(SrF \cite{Barry2014},
YO \cite{Collopy2018},
CaF \cite{Anderegg2017,Truppe2017,Williams2017}) and many others are being explored experimentally and theoretically \change{(AlF \cite{Truppe2019,Doppelbauer2021},
BaF \cite{Chen2017,Albrecht2020}, 
BaH \cite{Iwata2017}, 
Cs$_2$ \cite{Bahns1996}, 
MgF \cite{Xu2016}, 
RaF \cite{Isaev2010}, 
TiO \cite{Stuhl2008},
TlF \cite{Norrgard2017}, 
YbF \cite{Lim2018})}.
The list of laser cooled molecules is continuously growing with the recent addition of polyatomic species \change{
\cite{Isaev2016,Kozyryev2016b}
(CaOH \cite{Baum2020},
CaOCH$_3$ \cite{Mitra2020},
SrOH \cite{Kozyryev2017},
YbOH \cite{Augenbraun2020,Kozyryev2017a})}.
To enter the ultracold regime, subsequent cooling and trapping techniques, such as magnetic trapping, microwave trapping, dipole trapping and evaporative cooling of molecules, have been demonstrated \cite{McCarron2018,Williams2018,Wright2019,Anderegg2018,Anderegg2019,Son2020,Stuhl2012,NLeEDMCollaboration2021}.

 \change{Aluminum monochloride (}AlCl\change{)} has been proposed as an excellent candidate for laser cooling experiments \cite{Rosa2004,Yang2016,Wan2016}, as it is among a small set of molecules with very diagonal Franck-Condon factors, with theoretical estimates ranging from 99.88\%-99.93\% \cite{Langhoff1988,Ren2020,Yang2016,Wan2016}. Unlike typical molecules, this feature allows for the scattering of many photons on the electronic transition of the molecule before exciting it to higher vibrational states, where additional laser frequencies are needed to pump the molecule back into the cooling cycle.
To the best of our knowledge, the only other known diatomic molecule with higher predicted Franck-Condon factors is AlF \cite{Rosa2004,Truppe2019}.

A challenging aspect of any experiments with AlCl is that the transition wavelength is in the ultraviolet at 261.5\,nm.
\remove{
However, the availability of high-power solid-state laser systems at these wavelengths with a power of $>100$\,mW \cite{Mes2003b} in combination with its excellent photon scattering properties renders AlCl an ideal precursor for experiments that require a dense cloud of ultracold polar molecules with reasonable laser overhead.
Although our interest in AlCl is driven by its potential applications in AMO physics, 
this molecule is also of interest for the astrophysics and chemistry communities.}
\change{
However, high-power solid-state laser systems at these wavelengths with powers ranging from several hundred mW up to 1\,W have been realized \cite{Mes2003b,Ostroumov2008}.
The availability of such technology in combination with the excellent photon scattering properties render AlCl an ideal precursor to implement, for instance, quantum computing and simulation experiments or studies of dipolar quantum gases that require a dense cloud of ultracold polar molecules with reasonable laser overhead.
Furthermore, similar to the proposed schemes to produce cold fluorine \cite{Lane2012}, photodissociating cold and trapped AlCl could open up a path to producing cold chlorine atoms, whose transition frequencies are prohibitive for direct laser cooling and have eluded studies at low temperature so far.

AlCl is also of interest for the astrophysics and chemistry communities.}
For instance, it is among the major players in Al gas phase chemistry of cool stars. 
It has been predicted to exist and observed in the spectra of the circumstellar envelopes of carbon-rich stars, such as IRC+10216 \cite{Tsuji1973,Cernicharo1987,Agundez2012,Yousefi2018,Xu2020} and red asymptotic giant branch stars IK Tau and R Dor \cite{Decin2017}. Moreover, it is possible that the photosphere of the Sun contains AlCl in a detectable quantity since Al and Cl are found in the Sun's chemical composition \cite{Asplund2009}. AlCl may also be of interest to the spectroscopic study of exoplanets' atmosphere \cite{Tennyson2020,Wang2020}. From a chemical standpoint, AlCl has been suggested as a very efficient, cost-effective reduction agent to produce photovoltaic grade silicon \cite{Yasuda2009,Yasuda2011,Yasuda2011a}. Furthermore, the spectroscopic signature of AlCl has also been observed in rocket plumes \cite{McGregor1992,Oliver1992} and can be used to monitor chlorine content in potable water \cite{Parvinen1999}.

Spectroscopic measurements on AlCl over the past years have compiled lists of rovibrational transition frequencies \cite{Bhaduri1934,Holst1935,Mahanti1934,Sharma1951,Lide1965,Wyse1972,Hoeft1973,Ram1982,Rogowski1987,Mahieu1989,Mahieu1989a,Dearden1993,Ogilvie1994,Hedderich1993,Hensel1993,Saksena1998,Parvinen1998,Brites2008b}.
But many properties of AlCl, such as the electric dipole moment or chemical reaction rate coefficients, still remain unknown or have not been confirmed experimentally. \change{We note that the dipole moment of AlCl has been estimated theoretically to be $\approx\,1.59$\,D \cite{Yousefi2018}.}
In fact, astronomical models that rely on accurate knowledge of such parameters sometimes use substitute values of similar molecules to account for the missing parameters \cite{Ford2004,Agundez2012}. 
All the applications described above would benefit from a more detailed analysis of AlCl's spectroscopic properties and first-principles model of its electronic states.  

In this work, we produce and spectroscopically characterize a buffer-gas cooled sample of \remove{aluminum monochloride} \change{AlCl}.  We introduce the basic properties of AlCl, followed by an {\it ab initio} calculation of its potential energy curves for the $\Xstate$, $\Astate$ and $\apistate$ states.
Then the experimental setup to carry out absorption spectroscopy on the \XAtransition\ transition of a buffer-gas cooled sample of AlCl is presented. Finally, we discuss our experimental results on the equilibrium constants of the $\Astate$ state and our estimates for the Franck-Condon factors. The rigorous comparison of experiment and theory provides a solid foundation for future studies of its chemical properties, as well as for future laser-cooling experiments to generate ultracold samples where quantum effects become important.

\section{Laser Cooling Scheme of Aluminum Monochloride}

Aluminum monochloride is a closed-shell metal halide. The molecule has two stable isotopes, \alcl[27]{35} and \alcl[27]{37}, with natural abundances of $\approx 75.8\%$ and 24.2\%, respectively.
The three lowest energy levels of AlCl are the $\Xstate$ ground state, the $\apistate$ triplet and $\Astate$ singlet states.
In addition to the diatomic molecular vibrational and rotational excitations, the nuclear spin of Al ($I_\textrm{Al}=5/2$) and of both isotopes of Cl ($I_\textrm{Cl}=3/2$) exhibit in a hyperfine structure splitting of each state \cite{Hoeft1973,Hensel1993}.

The \XAtransition\ transition at 261.5\,nm has been proposed to be suitable for laser cooling AlCl \cite{Wan2016,Yang2016}. This transition offers a very high scattering rate due to the short radiative lifetime of the $\Astate$ state of 6.4\,ns \cite{Rogowski1987}.
While it is technically more challenging to produce high-power lasers in the ultraviolet, this short wavelength results in a \remove{slowdown} \change{change in velocity} of an AlCl molecule by about 2.5\,cm/s for each photon scattering event.
For a typical two-stage cryogenic-buffer gas beam of molecules with an average forward velocity of 60\,m/s \cite{Hutzler2012}, this means that only about 2,400 photons are required to slow such a beam of AlCl to a complete halt.
With the previous estimates for the Franck-Condon factors \remove{Cs} for the \XAtransitionvib{0}{0}\ transition of 99.88\% -- 99.93\% \cite{Wan2016,Yang2016} and the fact that the hyperfine structure is within the natural linewidth of the \XAtransition\ transition, a single-frequency cooling laser and one repumping laser to recover vibrationally excited molecules should be sufficient to slow a buffer-gas beam of AlCl below the typical capture velocity of a molecular magneto-optical trap of $\le 10$\,m/s.

As with other molecular cooling approaches, the combination of parity and dipole selection rules limits the excitation of AlCl to two rotational states \cite{Stuhl2008}. In the case of AlCl, the Q-transitions can be used for laser cooling since they are rotationally closed, similar to BH \cite{Hendricks2014} and AlF \cite{Truppe2019,Doppelbauer2021}.

Finally, the \Xapitransition\ singlet-triplet transition at around 406\,nm \cite{Herzberg2018,Ram1982,Mahieu1989} is dipole-forbidden and not suitable for laser cooling. Theoretical estimates of its natural lifetime are \remove{on the order of} $> 1$\,ms \cite{Wan2016}, which limits the photon scattering rate significantly.

\section{Ab Initio Calculations}
\label{sec:theory}

To compute the electronic states of the AlCl molecule, we use the MOLPRO 2015 quantum chemistry code \cite{Molpro2015} and multireference configuration interaction method with the Davidson correction (MRCI+Q). The basis set is aug-cc-pCVQZ (ACVQZ). The restricted Hartree-Fock (RHF) and the complete active space self-consistent field (CASSCF) calculations are carried out as prerequisites to the MRCI. The active space is the full valence active space.

While the two states of primary interest in the present study are the singlets $\Xstate$ and $\Astate$, we compute all states that result from the spin-orbit treatment of $\Xstate$, $\apistate$ and $\Astate$, similar to \cite{Wan2016}. The total number of $\Omega$-components is therefore 9: $\Xstate_{0^+}$, $\apistate_{0^-}$, $\apistate_{0^+}$, $\apistate_1$ (doubly degenerate), $\apistate_2$ (doubly degenerate) and $\Astate_1$ (doubly degenerate). The double degeneracy of the latter four components and a small splitting in the ``almost'' degenerate pair of $\apistate_{0^-}$ and $\apistate_{0^+}$ are features of diatomic molecules \cite{Herzberg1950}. The resulting number of distinguishable components is 6. 

\begin{figure}
    \centering
    \includegraphics[width=\linewidth]{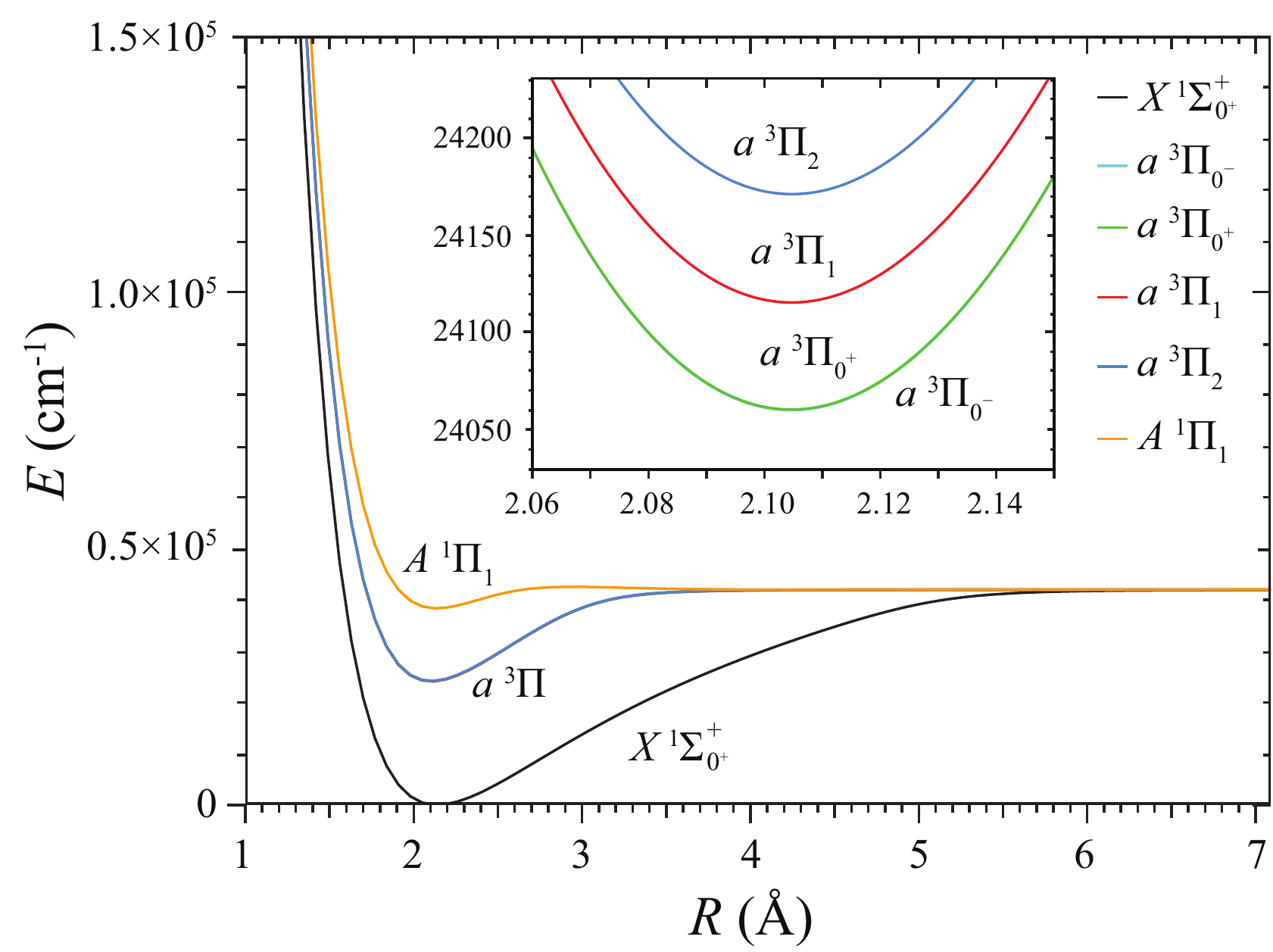}
    \caption{Potential energy curves of AlCl as a function of the interatomic distance $R$. There are two singlet states $\Xstate$ (black) and $\Astate$ (yellow) and one intermediate triplet state $\apistate$. An inset shows four $\Omega$-components of the triplet state: $\apistate_{0^-}$ (light blue), $\apistate_{0^+}$ (green), $\apistate_1$ (red) and $\apistate_2$ (blue). The splitting between the two $+/-$ sub-components of $\apistate_0$ is very small and is not visible.}
    \label{fig:pes}
\end{figure}

\rfig{fig:pes} shows six potential energy curves for AlCl computed in this work. As one can see, all curves have the same shape and all their minima are located at nearly the same position along the interatomic distance $R$. Thus, the AlCl molecule will have good Frank-Condon factors between these states and is therefore a strong candidate for laser cooling. The multiple components of the $\apistate$ states are indistinguishable at the large scale, but show the spin-orbit splittings when zoomed in (see inset). The figure closely resembles Figure~2 in \cite{Wan2016}.

The optimal bond lengths $R_e$ and energies $T_e$ for each electronic state are given in \rtab{tab:re-and-te}. The computed $R_e$ values are very close to that of the previous theoretical study \cite{Wan2016} and our theoretical predictions for $T_e$ also match the old ones quite well. For the intermediate state $\apistate$, our $T_e$ values are located in between \cite{Wan2016} and experiments \cite{Sharma1951,Ram1982}. However, both theoretical studies, this work and \cite{Wan2016}, underestimate $T_e$ for that state by 500 to 900\,cm$^{-1}$ on average. In contrast, the excited state $\Astate$ is reproduced much more accurately. This work overestimates its $T_e$ by 65.08\,cm$^{-1}$, whereas the previous study \cite{Wan2016} underestimates it by 29.56\,cm$^{-1}$.

\begin{table*}
    \caption{Equilibrium distances $R_e$ (\AA) and electronic energies $T_e$ (cm$^{-1}$) for AlCl.}
    \begin{ruledtabular}
    \begin{tabular}{@{\extracolsep{4pt}}ccccccccccc@{}} 
        \multirow{2}{*}{State} &
        \multicolumn{3}{c}{$R_e$} &
        \multicolumn{4}{c}{$T_e$} &
        \multicolumn{3}{c}{$\Delta T_e$} \\
        \cline{2-4} \cline{5-8} \cline{9-11}
        & This work & Ref.~\cite{Wan2016} & $\Delta$ &
          This work & Ref.~\cite{Wan2016} & Exp.~\cite{Sharma1951} & Exp.~\cite{Ram1982} &
                      Ref.~\cite{Wan2016} & Exp.~\cite{Sharma1951} & Exp.~\cite{Ram1982} \\
                      \hline
        $\Xstate_{0^+}$        & 2.1373 & 2.1374 &  0.0001 &        0 &        0 &                      0 &                        0 &       0 &                       0 &                       0 \\
        $\apistate_{0^-}$      & 2.1044 & 2.1049 &  0.0005 & 24060.33 & 23905.13 & \multirow{2}{*}{24528} & \multirow{2}{*}{24793.1} & -155.20 & \multirow{2}{*}{467.52} & \multirow{2}{*}{732.62} \\
        $\apistate_{0^+}$      & 2.1044 & 2.1049 &  0.0005 & 24060.62 & 23905.41 &                        &                          & -155.21 &                         &                         \\
        $\apistate_1$          & 2.1044 & 2.1050 &  0.0006 & 24115.36 & 23959.97 &               24593.84 &                 24855.46 & -155.39 &                  478.48 &                  740.10 \\
        $\apistate_2$          & 2.1045 & 2.1050 &  0.0005 & 24170.76 & 24015.17 &                  24658 &                 24919.75 & -155.59 &                  487.24 &                  748.99 \\
        $\Astate_1$            & 2.1340 & 2.1330 & -0.0010 & 38319.08 & 38224.44 &                      - &                    38254 &  -94.64 &                       - &                  -65.08 
    \end{tabular}
    \end{ruledtabular}
    \label{tab:re-and-te}
\end{table*}

\change{
We note that our level of ab initio theory is identical to that of the previous theoretical study on AlCl \cite{Wan2016}. However, we also performed an extensive set of calculations which go beyond the current state-of-the-art in an attempt to improve the accuracy, as will be discussed in more detail below. The basis set used in \cite{Wan2016} has the letter ``C'' in the name, which means that the basis was optimized to include core-valence or core correlation effects. This suggests the usage of all-electron treatment for the problem. However, there was no mention of ``unfreezing'' the outer core or inner-shell orbitals in that study, and all of their calculations seem to correlate valence electrons only, which is the default in a typical electronic structure code. Using an all-electron basis set (another name for a core correlated basis set) does not enable correlations beyond the valence shell automatically.

The specialized basis set used in \cite{Wan2016} motivated us to pursue an all-electron calculation. In this high-level calculation, all core orbitals were unfrozen and correlated except the two very low-lying 1\textit{s} orbitals, according to the definition of ``all-electron'' \cite{Peterson2002}. One of these two still-frozen orbitals corresponds to the 1\textit{s} orbital of aluminum, while another to the 1\textit{s} of the chlorine atom. Using this higher-level theory, we found that the spin-orbit splittings between the components of the intermediate $\apistate$ states improved. For example, an all-electron treatment gives $A^{SO} (\apistate_1-\apistate_0) = 63.20$ cm$^{-1}$ and $A^{SO} (\apistate_2-\apistate_1) = 64.44$ cm$^{-1}$, which are in excellent agreement with the experimental values 65.84 and 64.16 cm$^{-1}$ \cite{Sharma1951}, or 62.36 and 64.29 cm$^{-1}$ \cite{Ram1982}. Without an all-electron treatment the theoretical splittings are 54.74 and 55.39 cm$^{-1}$, respectively. Using the higher-level theory, we also found that the $T_e$ values for the $\apistate$ components improved as well, by a few hundred wave numbers.

However, we also discovered that other quantities disagreed with the experiments when using the higher-level theory. Namely, the $T_e$ value of the $\Astate$ state and the transition energies involving vibrational excitations $Q_{vv'}$. In particular, the $T_e$ of the $\Astate$ became more than a thousand wave numbers larger than the experimental one, whereas without the all-electron treatment the $T_e$ error is only 65.08 cm$^{-1}$. Using a manually-corrected $T_e$, so that the theoretically-predicted energy for the \XAtransitionvib{0}{0} transition matches the experimental one, the errors between the high-level theory and experiment \cite{Ram1982} for the $Q_{11}$, $Q_{22}$, $Q_{33}$ and $Q_{44}$ transitions were -22.26, -39.39, -52.70, -59.76 cm$^{-1}$, respectively. In contrast, performing a similar treatment but \textit{without} the core correlation, we found that the errors between the lower-level theory and experiment were much smaller:  0.29, 2.80, 6.23 and 13.01 cm$^{-1}$, respectively.

In summary, an all-electron treatment (higher level theory) was found to improve the $\apistate$ state, but worsen the $\Astate$ state. As discussed in detail in \cite{Puzzarini2013}, the choice of ab initio theory as well as additional manual shifts in $T_e$ are often required when comparing experimental spectroscopy to first principles based theory. Since the focus of the present study is the \XAtransitionvib{0}{0} transition, we therefore decided not to use the all-electron treatment. We also checked that switching from the specialized ACVQZ basis set to AVQZ lowers the $T_e$ for both the $\apistate$ and $\Astate$ states by 15 cm$^{-1}$. Thus, there is little difference between these two basis sets and either one can be used reliably. Ultimately, we chose the ACVQZ to enable a direct comparison with the results of \cite{Wan2016}, which renders both theoretical studies identical (except that we use a newer version of MOLPRO). We also note that another very similar study \cite{Yang2016} used the standard aug-cc-pVQZ (AVQZ) basis set.
}

In preparation for computing the one-dimensional wave functions for AlCl, we fit the {\it ab initio} curves with a standard Morse potential
\begin{equation}\label{eq:morse}
  V(R) = D(e^{-\beta(R-R_e)}-1)^2,
\end{equation}
where $D$ is the potential depth, $R_e$ is the equilibrium bond distance and $\beta$ is the inverse of the potential width. We fit all curves simultaneously using a simplex-based minimization algorithm (AMOEBA code) \cite{Numrec1986} with custom weights, so that the fitting algorithm prioritizes low energy points over high energy ones. Such weighting is important, because we are interested in only a few quanta of vibrational excitation in each electronic state. Thus, there is no need for high accuracy fitting in the high energy (or high temperature) region. This approach also allows us to use the standard Morse form even for the excited state $\Astate$ with a barrier, which is barely visible in \rfig{fig:pes}. The standard Morse expression does not reproduce the barrier. However, both the barrier top at 4210\,cm$^{-1}$ and the asymptotic value of the potential 3683\,cm$^{-1}$ (at 7\,\AA, relative to the minimum energy) are beyond the fitting energy window of 1000\,cm$^{-1}$, that covers the $v=0$, 1 and 2 vibrational levels of $\Astate$. The fitting window for the other states is 2000\,cm$^{-1}$, and the total fitting error is 9.87\,cm$^{-1}$ (i.e., six times smaller than the $T_e$ error). To further reduce the fitting error, an expression with more than three parameters will be needed.

After fitting the {\it ab initio} potential curves to the Morse expression \req{eq:morse}, we can perform a numerically exact solution of the one-dimensional diatomic ro-vibrational Schr\"odinger equation (see Eq.~152 in Ref. \cite{Pack1987}). This approach is more general than a standard harmonic oscillator expansion about the equilibrium geometry and includes anharmonic contributions. A Numerov propagator was used to compute the vibrational wave functions and energies for each rotational state \cite{Johnson1977,Johnson1978}. A uniform grid in $R$ was used which consisted of $6000$ points spanning the range $R_i=1.5$ $a_0$ to $R_f=20.0$ $a_0$ inclusive. The vibrational wave functions were used to compute the appropriate overlaps or Franck-Condon factors (see \rtab{tab:FCs}). The rotational wave functions are analytic and expressed in terms of standard spherical harmonics \cite{Pack1987}. We note that this same approach was used in our recent calculations for the diatomic NO molecule in the Ar-NO collision system \cite{Teplukhin2020}. For singlet AlCl the total angular momentum operator ${\hat J}$ is the sum of the diatomic rotation operator ${\hat R}$ and the electronic angular momentum ${\hat L}$, ${\hat J} = {\hat R} + {\hat L}$. Thus, the rotational term in the diatomic Hamiltonian is proportional to ${\hat R}^2 = ({\hat J} - {\hat L})^2$. Ignoring off-diagonal couplings to other electronic states and $\Lambda$ doubling, this expression can be simplified to ${\hat R}^2= {\hat J}^2 - {\hat L}_z^2$ where $L_z$ is the component of electronic angular momentum along the internuclear axis. The rotational energy eigenvalues are $E_J = B\,[J(J+1) - \Lambda^2]$ where $B$ is the rotational constant and $\Lambda=0$ and $1$ for the $^1\Sigma$ and $^1\Pi$ states, respectively. Thus, the total angular momentum quantum number $J$ is an integer and for the $^1\Sigma$ state it starts at zero ($J=0,1,2,\ldots$) whereas for the $^1\Pi$ state it starts at one ($J=1,2,3,\ldots$). The numerically computed {\it ab initio} based ro-vibrational energies for the $v=0$ and $1$ manifolds of the $^1\Sigma$ and $^1\Pi$ electronic states were used to compute the $Q$, $R$, and $P$ transition frequencies reported in \rfig{fig:spectrum_00}, \rtab{tab:freqs35} and \rtab{tab:freqs37}. Excellent agreement is seen between the {\it ab initio} based frequencies and experiment which is discussed in more detail in \rsec{sec:exp_results}. The only adjustments to the {\it ab initio} energies were the following: (1) the {\it ab initio} computed $T_e$  value reported in \rtab{tab:re-and-te} was shifted slightly (by $\Delta T_e = - 66.05236$ cm$^{-1}$) to match the experimentally observed transition frequency for $Q_{00}$; and (2) an additional energy shift of ($\delta E = - 0.78831$ cm$^{-1}$) was added to the $v=1$ manifold to match the experimental transition frequency for $Q_{11}$ (the $\delta E$ is due to additional small vibrational anharmonicities not included in the Morse potential).

\section{Experimental Setup}

\begin{figure}[ht]
    \centering
    \includegraphics[scale=0.44]{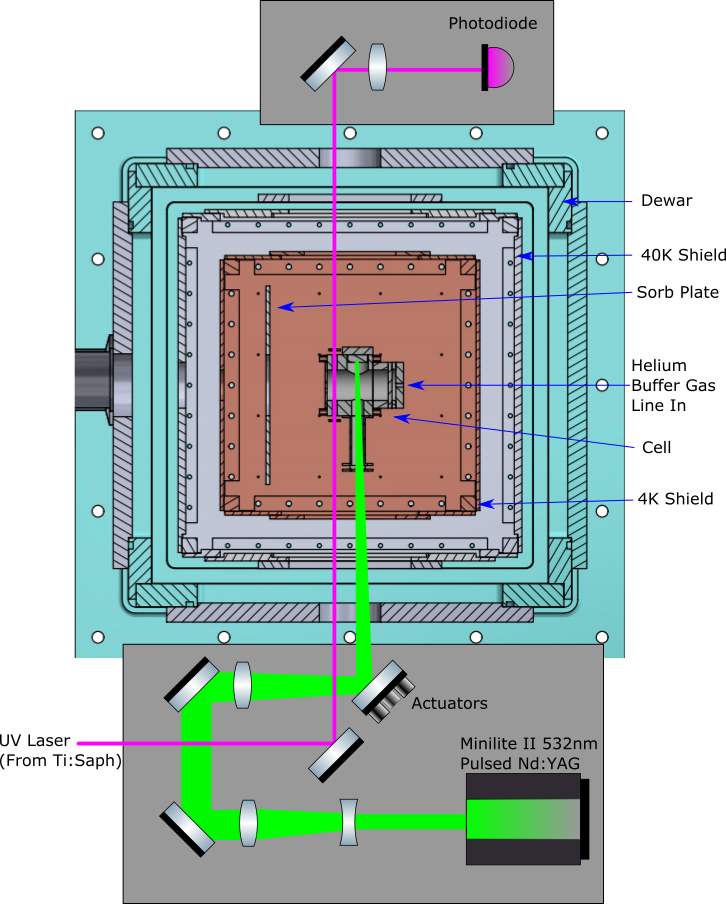}
    \caption{Horizontal-plane cross section of the CBGB, depicting the exterior, room-temperature Dewar (teal), the 40\,K aluminum shields (light grey), the 4\,K copper shields (orange), and the central copper cell (dark grey). A copper sorb plate is shown (dark grey). Also shown is the optical setup of the ablation laser (green), and the UV spectroscopy laser (magenta) passing through the cell.}
    \label{fig:apparatus}
\end{figure}

Our experimental apparatus
\change{(see \rfig{fig:apparatus})}
has two main components: a cryogenic buffer-gas beam source (CBGB) \cite{Hutzler2012} to generate the AlCl gas, and a frequency-tripled Ti:Sapphire CW laser system for spectroscopy \cite{Mes2003b}.

\subsection{Buffer-Gas Beam Source}
\label{sec:expt_setup}

At the heart of our setup is a cryogenic helium buffer-gas beam source to hold the sample. 
The cell is cooled down with a two-stage pulse-tube cooler (Cryomech PT420). To minimize the heat load, the cell is surrounded by two layers of heat shields (Cu 101-OFE and Al 6061).
The first stage of the pulse tube is connected to the Al heat shield, cooling it to about 40\,K, whereas the second is connected to both the cell and the copper heat shield.
The aluminum shield is covered in a few layers of multi-layer insulation (Lakeshore NRC-2) to further reduce heat loads on the system. 
The inside of the copper shield walls and bottom are coated with coconut charcoal (sorbs) to provide effective cryogenic pumping for He buffer-gas at low temperatures \cite{Sedgley1987,Tobin1987}. These sorbs are attached using an epoxy consisting of a 100:7.5 ratio of Stycast 2850FT (Loctite) to catalyst 23LV (Loctite) by weight. One additional sorb plate is mounted inside the copper shields.

Optical access is provided by glass viewports at the outer vacuum chamber and through holes in the heat shields. The latter was left open to minimize absorption of the transmitted UV spectroscopy light in glass. The additional thermal heat load results in a steady-state temperature of the buffer-gas cell of about 4.5\,K, which is slightly higher than the specified base temperature of the pulse tube without heat load of 2.8\,K.

The cell contains a solid precursor target, which is glued with Stycast 2850FT on a copper piece and mounted inside the cell. \change{The outer dimensions of the cell are $3.8 \times 3.8 \times 6.5$\,cm, with an internal volume of $\approx$28 cm$^3$ and an exit aperture of 5\,mm.}
A fill line, which is thermally anchored to the heat shields and connected to the cell, flows cold helium buffer gas into the cell at a rate of 3--4\,sccm. The initial part of the fill line with an outer-diameter of 1/8'' from the outside of the vacuum chamber to the Al heat shield is made of stainless steel and the remaining part is made of copper. During the experiments, helium buffer gas flows continuously into the cell and out again through the exit aperture of the cell.

\begin{figure}[ht]
    \centering
    \includegraphics[scale=0.75]{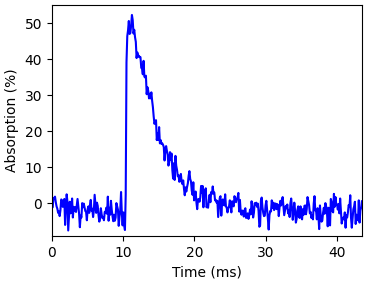}
    \caption{Absorption transient of \alcl{35} on the \XAtransition\ transition. The ablation laser fires at 10\,ms.}
    \label{fig:timetrace}
\end{figure}

To perform absorption spectroscopy on AlCl, the solid precursor is ablated with a short-pulsed laser (532 nm, 5$\pm$2 ns pulse Continuum Nd:Yag, $\approx 10$\,mJ) and the transmission of a tunable spectroscopy UV laser beam is monitored with an amplified photodiode (Thorlabs PDA25K2).
The collimated spectroscopy beam \remove{passes} \change{enters} the cell through 3.0\,mm thick glass windows with a UV anti-reflective coating. The ablation laser is focused into the cell through a glass window that is offset from the target to avoid dust accumulation of the ablation window. 
\change{
The diameter of the ablation laser is $\approx\,80\mu$m on the target.
}
Upon ablation, collisions with the Helium buffer-gas cool down the ablation plume, which results in a cold sample of AlCl in the gas phase. 
This cooled sample is carried by the helium flow through the UV spectroscopy beam 1.9\,cm downstream from the ablation point.
An example of an absorption transient is shown in \change{\rfig{fig:timetrace}}.
The experimental sequence is repeated with a rate of $\approx 1$\,Hz to allow the cell to cool down again in between ablation shots.
To avoid drilling a hole in the target with the ablation process, the last mirror of the ablation laser outside the vacuum chamber is continuously steered using actuators on both mirror axes (Conex CC, New Focus) to raster over the target.

\subsection{Laser Setup}

We probe the \XAtransition\ transition in AlCl with 261.5\,nm light developed from frequency tripling a CW Ti:Sapphire (Ti:Saph) laser.
The Ti:Saph is pumped by a 532\,nm Nd:Yag (Sprout, Lighthouse Photonics) and frequency stabilized to the readout of a wavemeter (High Finesse, WS-7) using a software proportional-integral controller (PID).

\begin{figure*}[ht]
    \centering
    \includegraphics[width=0.8\linewidth]{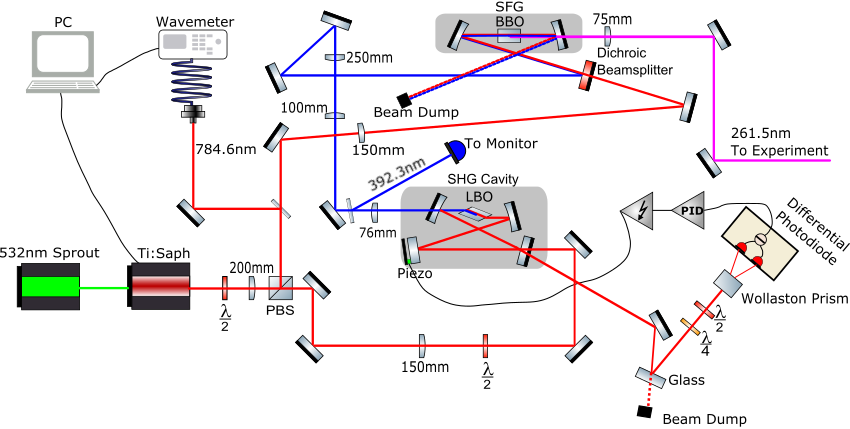}
    \caption{The output of a CW Ti:Saph is frequency doubled with a LBO crystal enhancement cavity to create 392.3\,nm light. The 392.3\,nm is combined with fundamental 784.6\,nm light and focused single pass into a BBO crystal to produce 261.5\,nm UV light used to perform absorption spectroscopy. The enhancement cavity is stabilized with an electronic Red Pitaya PID. The wavemeter measurements are calibrated using a Rb atomic reference, and the Ti:Saph is controlled and stabilized with a software \change{PID} and \change{an} \remove{a}\change{A}rduino \change{is used to supply the control signal to the laser} \remove{PID}.}
    \label{fig:lasers}
\end{figure*}

To calibrate our frequency measurements, we use a two-fold approach: First, we calibrate the wavemeter's frequency output by comparing the output to a Doppler-free saturated absorption spectroscopy of Rubidium with another Ti:Sapph laser (Coherent 899-21). Second, we monitor the wavemeter drifts by comparing it with a Helium-Neon laser (NewFocus). Both methods are applied before the absorption spectroscopy scans.
Using this setup, we estimate an upper limit on the average error of each frequency measurement of $\approx 15$\,MHz, where the main source of this error is the slow feedback loop of the software lock that compensates for long-term drifts of the frequency of the Ti:Saph. We use these estimates to do a Monte Carlo simulation to determine the errors of the Dunham coefficients.

The output of the Ti:Saph is focused into a lithium triborate (LBO) nonlinear crystal (Newlight Photonics, angle cut at $\theta = 90^\circ, \phi = 33.2^\circ$, brewster cut for 784.6\,nm) enhancement cavity. We create $\approx 100$\,mW of 392.3\,nm light and utilize a bow-tie geometry so that high laser intensities will not create a standing wave and damage the crystal. The SHG cavity is length-stabilized to match the Ti:Saph frequency using a H\"ansch-Couillaud locking scheme by feeding back onto a piezo-actuated mirror. The feedback loop is controlled by an electronic PID controller (Red Pitaya).

The 392.3\,nm light from the SHG cavity is then combined in a sum-frequ\change{e}ncy-generation (SFG) process with the fundamental 784.6\,nm light and focused single pass into a barium borate (BBO) nonlinear crystal (Newlight Photonics, angle cut at $\theta = 45.5^\circ$, coated with  a broadband anti-reflection coating to cover all three wavelengths). The SFG process produces $\approx 30\,\mu$W of 261.5\,nm UV light, which is then directed through the buffer-gas cell to perform absorption spectroscopy.

\subsection{Target Preparation}
\label{sec:target_prep}

Two different chemical precursors were used for the absorption spectroscopy, AlCl$_3$ and KCl+Al.  For the AlCl$_3$ pellet, 1\,g of 98\% sublimed, anhydrous aluminum trichloride (Sigma Aldrich) was used.  For the KCl+Al pellet, 0.9\,g of 99\% BioXtra potassium chloride (Sigma Aldrich) was mixed with 0.1\,g of 99.95\% aluminum powder ($<75$ micron particle diameter) obtained from Sigma Aldrich. 
Each of the powder mixtures was put into a 12\,mm pellet die and pelletized in a hydraulic press with 6000\,psi for 1\,minute. Each pellet weighed about 1\,g and was 12\,mm in diameter and 3.5\,mm in height. The samples were then chiseled in half to fit on the copper holder that would be loaded into the chamber, making each target about 0.5\,g. They were glued with Stycast to a copper holder. A thin layer of epoxy was spread onto the copper holder and the pellet was placed on top. The entire copper holder was then wrapped in parafilm to avoid exposure to the atmosphere. After 5\,hours of drying, the copper holder was transferred into the vacuum chamber. The total duration of air exposure during sample preparation was less than 60\,minutes.

The AlCl$_3$ target was used for early observations, and initially produced an observable amount AlCl. This precursor target was used to obtain the spectroscopy on the \XAtransitionvib{0}{0} transition.
For the later observations of weaker lines, such as the P branch of the \XAtransitionvib{1}{1} manifold, we used a mixture of KCl and Al for a pellet.
The main reason for switching the precursor was that the pressed AlCl$_3$ targets turned out to be less reliable in the production of AlCl.
We attribute this to the quick degradation of the AlCl$_3$ targets under atmospheric exposure, where HCl is formed in the reaction with the water content in air.
\change{A detailed, comparative study on the yields of mixture targets with different salts is in preparation by our group.}


\section{Experimental Results}
\label{sec:exp_results}

We obtained spectra for the \XAtransitionvib{0}{0} and the \XAtransitionvib{1}{1} transitions. 
For the 0-0 vibrational band, we observed the P, Q and R transitions for both the \alcl{35} and \alcl{37} isotopologues, as shown in \rfig{fig:spectrum_00}. We average ten ablation shots for each frequency step to acquire the spectrum.
In the case of the 1-1 band, we were able to observe the Q and R transitions for \alcl{35}, and the Q and one R transition for \alcl{37}, with the same procedure, as shown in \rfig{fig:spectrum_00}. 
The corresponding P transitions, however, appeared to be much weaker in line strength and required 100 averages to provide sufficient signal-to-noise.

\change{The rovibrational energies of the $\Xstate$ and $\Astate$ states can be described with a Dunham type model \cite{Dunham1932}}
\begin{eqnarray}
\label{eq:E_dun}
\change{E(v,J) = \sum_{k,l}Y_{kl} (v + 1/2)^k [J(J+1)-\Lambda^2]^l}
\end{eqnarray}
\change{with $\Lambda = 0\,(\Lambda = 1)$ for the $\Xstate\, (\Astate)$ state, where $Y_{kl}$ are the Dunham coefficients.}
To determine the line centers, we fit a Voigt profile to each of the measured transitions to account for Doppler and pressure broadening due to the finite temperature and the He buffer gas in the cell. 
The resulting line centers are presented in \rtab{tab:freqs35} and \rtab{tab:freqs37}.
These transition frequencies were used to do a least-square fit to determine the Dunham coefficients of the $\Astate$ state, as described in detail in \rsec{sec:molecular_constants}. For the $\Xstate$ state, the values from a previous high-resolution infrared spectroscopy studies on AlCl \cite{Hedderich1993} were used.

\subsection{Rotational Temperature}

\begin{figure}
    \centering
    \includegraphics[scale=0.7]{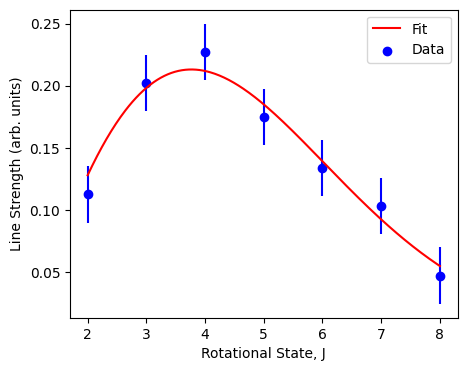}
    \caption{Measured rotational line intensities at different rotational quantum states of the P branch for $v=0$ of Al$^{35}$Cl in the buffer-gas cell, integrated from 0.7 to 3.1\,ms, after ablation. The fit corresponds to a rotational temperature of $8.3(4)$\,K.}
    \label{fig:rot_population}
\end{figure}

\begin{figure*}[ht]
    \centering
    \includegraphics[width=0.99\linewidth]{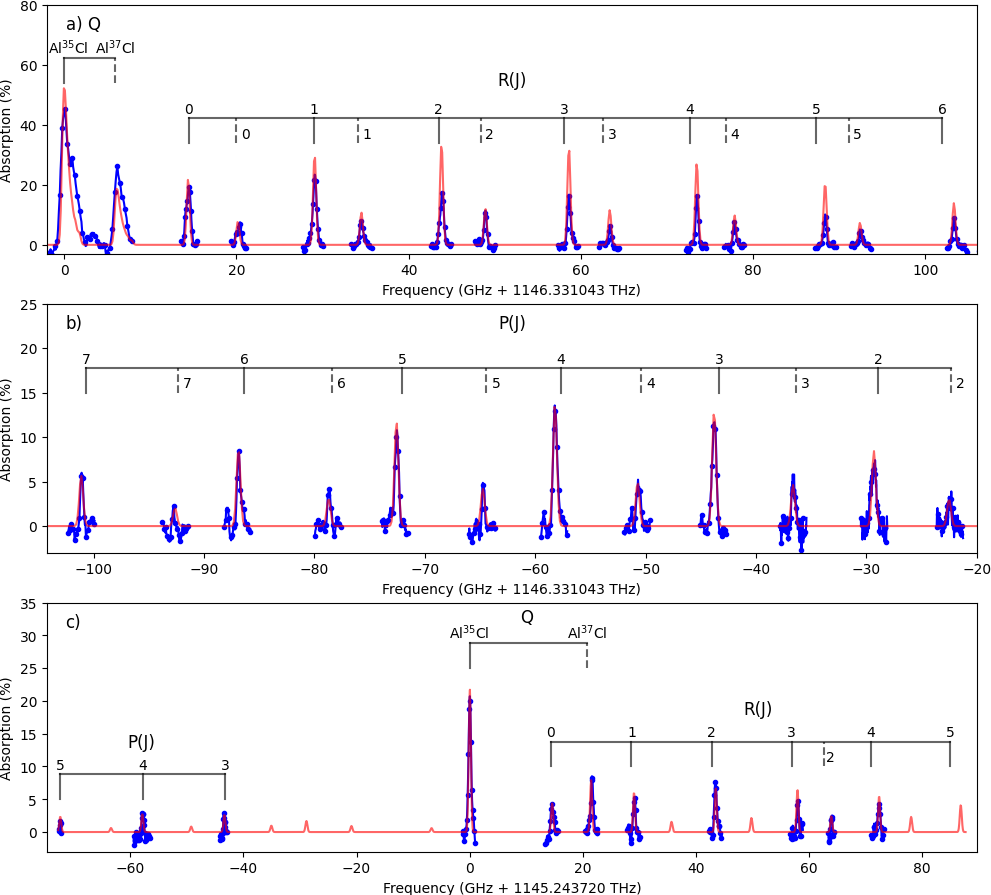}
    \caption{
    \label{fig:spectrum_00}
    {\bf (a)} Q branch, R branch, {\bf (b)} P branch absorption spectrum of the \XAtransitionvib{0}{0} transition.
    {\bf (c)} Q, R, and P branch of the \XAtransitionvib{1}{1} transition.
    Blue dots are the measured data. Red solid lines are the fit to the mass-reduced Dunham model.
    A moving average of $\pm 3$ is applied to all data points for clarity. Black vertical lines are frequencies predicted by the {\it ab initio} theory for \alcl{35} (solid) and \alcl{37} (dashed) and are labeled with the rotational quantum number $J$ of the $\Xstate$ state.
    }
    
\end{figure*}

We extract the rotational temperature of the ablation AlCl gas from the absorption spectrum of the P branch transitions of \alcl{35} in the $v = 0$ vibrational manifold. 
\rfig{fig:rot_population} shows the measured absorption, integrated from 0.7 to 3.1\,ms, after the ablation pulse as a function of the rotational quantum number $J$ in the $\Xstate$ state.

Assuming a Boltzmann distribution for the rotational state distribution,
\begin{equation}
\label{eq:rot_population}
    N_J \propto (2J + 1) e^{- E_J / (\kB T)}\quad,
\end{equation}
\change{where $E_J = B_v J (J+1)$ and $B_v \approx Y_{01} + Y_{11} (v + 1/2)$ is the rotational constant,}
and taking the $J$-dependent transition strengths into account with the H\"onl-London factors \cite{Hansson2005},
\begin{equation}
\label{eq:hoenl_london}
    S_{J} = \frac{(J - \Lambda - 1)(J - \Lambda)}{2J} \quad,
\end{equation}
where $\Lambda = 0$, a least-square fit to the measured rotational line strengths to
$I(J, T) \propto S_{J} \cdot N_J(T)$ yields a rotational temperature of 8.3(4)\,K. 
The elevated temperature in comparison to the base temperature of the cell is attributed to the additional heat load from each ablation shot and is measured only a few ms after ablation before the gas of AlCl molecules have completely thermalized with the helium buffer gas.

\subsection{Molecular Constants}
\label{sec:molecular_constants}


\remove{The rovibrational energies of the $\Xstate$ and $\Astate$ states can be described with a Dunham type model \cite{Dunham1932}}
\remove{with $\Lambda = 0\,(\Lambda = 1)$ for the $\Xstate\, (\Astate)$ state.}

Using a least-square fit to the measured line centers, we find excellent agreement of the data with \change{the Dunham model (\req{eq:E_dun})} using a separate set of Dunham coefficients for each isotopologue, $Y^{35}_{kl}$ and $Y^{37}_{kl}$. For the fitting procedure, we vary only the coefficients for the $\Astate$ state, whereas the $\Xstate$ state coefficients were taken from previous high resolution measurements \cite{Hedderich1993}. In addition, we restricted the number of coefficients to a minimum, such that any systematic offsets of data and measured line centers are reduced to a minimum. The resulting sets of Dunham coefficients for both isotopologues are given in \rtab{tab:Ys_fitted}, along with the average difference between the prediction and the measured line centers (mean line error) of 26 and 41\,MHz, for each respective isotope. We note that our current signal-to-noise renders it very challenging to observe more rotational states of the \alcl{37} isotope in the $v=1$ manifold, which in turn limits our ability to determine an accurate value for the $Y_{11}^{37}$ coefficient.

\begin{table}[h!]
    \centering
     \caption{\label{tab:Ys_fitted}Dunham coefficients in units of cm$^{-1}$ for the $\Astate$ state obtained from our measured line centers.}
     \begin{ruledtabular}
    \begin{tabular}{c|rr}
         & \alcl{35} & \alcl{37} \\
        \hline
        $Y_{00}$ & 38257.4210(4) & 38257.3401(7) \\ 
        $Y_{10}$ & 441.3320(6) & 436.2127(6) \\
        $Y_{01}$ & 0.24534(2) & 0.23833(2) \\
        $Y_{11}$ & -0.00265(3) & \\
        \hline
        Mean Line Error & 26 MHz & 41 MHz
    \end{tabular}
      \end{ruledtabular}
\end{table}

On the other hand, using a unified fitting model that describes both isotopologues simultaneously allows for extracting additional information since the isotope dependence of rotational and vibrational constants adds additional restrictions on the fit.
This mass-reduced approach takes the isotope dependence of $Y_{kl}$ into account by scaling each coefficient with a factor $\mu^{-(2k+l)/2}$, where $\mu$ is the reduced mass of AlCl \cite{Watson1980}.

Following this procedure, we arrive at a set of $U_{kl}$ listed in \rtab{tab:Us_fitted}. While this model offers the advantage
that the energies of the rovibrational quantum states of all isotopes can be predicted with a single set of mass-reduced Dunham coefficients, in the case of AlCl, it results in an average line error which is higher by a factor of two in comparison to the approach with two separate sets of Dunham coefficients.
In addition, we encountered systematic frequency differences between the data of the two isotopes and the mass-reduced fitting model.

\begin{table}[h]
    \centering
     \caption{\label{tab:Us_fitted}Mass-reduced Dunham coefficients in units of cm$^{-1}$ for the $\Astate$ state obtained from our measured line centers. The third column are the Dunham coefficients when including one Born-Oppenheimer breakdown correction factor ($\Delta_{00}^\textrm{Cl}$).}
     \begin{ruledtabular}
    \begin{tabular}{c|r|r}
        $U_{00}$ & 38253.33(2) & 38253.31(2) \\ 
        $U_{10}$ & 1764.9(2) & 1766.1(2) \\
        $U_{20}$ & -83.0(4) & -85.4(4) \\ 
        $U_{01}$ & 3.7377(3) & 3.7367(3) \\
        $U_{11}$ & -0.165(2) & -0.157(2) \\
        $\Delta^\textrm{Cl}_{00}$ & & -0.158(7)\\
        \hline
        Mean Line Error & 72 MHz & 31 MHz
    \end{tabular}
      \end{ruledtabular}
\end{table}

We find that a more advanced mass-reduced model which takes the effects of rovibronic interactions between electronic states into account results in better agreement of model and data. This model modifies the reduced mass in the Dunham coefficients by introducing isotope-dependent Born-Oppenheimer breakdown factors \cite{Watson1973,Watson1980},
\begin{eqnarray}
\label{eq:U_kl}
Y_{kl} = \mu^{-(k+2l)/2} \left( 1 + \frac{m_e}{m_\textrm{Cl}} \Delta^\textrm{Cl}_{kl} \right) U_{kl}\quad.
\end{eqnarray}

Using this approach while introducing a single breakdown coefficient $\Delta_{00}^\textrm{Cl}$, we find excellent agreement of the data and the fit model with the $U$ coefficients presented in the third column of \rtab{tab:Us_fitted}.
The comparison of our model and the data is also shown in \rfig{fig:spectrum_00}.

We note that introducing the $\Delta^\textrm{Cl}_{00}$ correction term is ambiguous, given the number of transitions we were able to observe with our system, and a similar agreement of data and fit model can be achieved using $\Delta^\textrm{Cl}_{10}$.
To our knowledge, there are no estimates available in the literature for these correction factors for the $\Astate$ state and only one measurement is available for the $\Xstate$ state \cite{Hedderich1993}.
The fitted value of $\Delta_{00}^\textrm{Cl} = -0.158$\,cm$^{-1}$ leads to shifts of $T_e$ of order $\approx 0.1$\,cm$^{-1}$ and is of order unity, as expected and discussed in \cite{Watson1980} for electronic states that are well-isolated.

Finally, the unexpected relative intensities of the P and R branches of the 1-1 manifold that could be an artifact of the preparation method of AlCl.
A possible explanation is that a part of the AlCl sample is in the excited $\Astate$ state during the laser absorption process. This population, in turn, results in competing absorption and emission processes, \XAtransition\ and \AXtransition, each of which have different H\"onl-London factors, which leads to weaker P transitions \cite{Hansson2005}.
We assume two possible pathways that could lead to a significant excited state population: AlCl molecules are created in electronic states with very high energies during the ablation process, e.g.~the triplet $c^3\Sigma^+$ state \cite{Langhoff1988}, which decay via slow spontaneous emission or collisional dexcitation into the $\Astate$, or AlCl molecules are formed in the $\Astate$ directly via collisions in the ablation plume after ablation. To test the first hypothesis, we examined the dependence of the line strength on the power of the ablation laser pulse, but no significant change in line strength could be observed. We will leave a more detailed analysis of this phenomenon to a future study.

\subsection{Franck-Condon Factors}

\begin{table*}[ht!]
    \caption{\label{tab:FCs}Franck-Condon factors estimates obtained for the \XAtransition\ transition from the harmonic oscillator approximation using the experimental rotational constant compared to estimates from {\it ab initio} theory.}
    \begin{ruledtabular}
\begin{tabular}{lllll}
    $f_{00}$ & $f_{01}$ & $f_{11}$ & $f_{10}$ & Ref.\\
    \hline\\
    0.9988 & 0.0012 & 0.9961 & 0.0019 & This work (\alcl{35})\\
    0.9989 & 0.0011 & 0.9961 & 0.0019  & This work (\alcl{37})\\
    0.9988 & 0.0003 & 0.9965 & 0.0003 & This work (theory)\\
    0.9988 & 0.0005 & 0.9970 & 0.0005 & \cite{Wan2016}\\
    0.9993 & $0.1157 \times 10^{-8}$ & 0.9960 & $0.9677 \times 10^{-6}$ & \cite{Yang2016}
    \end{tabular}
    \end{ruledtabular}
    \label{tab:fcf_all}
\end{table*}

We estimate the Franck-Condon factors (FCF) for the \XAtransition\ transition by approximating the potential energy as a harmonic oscillator. The FCF is then given by
\begin{equation}
    f_{v'v} = \bra{\psi_{v'}} \ket{\psi_v}\quad,
\end{equation}
where $\psi_{v'} (\psi_{v})$ is the simple  harmonic oscillator vibrational wave function of the $\Xstate (\Astate)$ state.
The wave functions depend on the molecular bond length $R_e$ and the vibrational constant $\omega_v$. The equilibrium bond length is determined by the rotational constant and arises from the solution of the rigid rotor potential 
\begin{equation}
\label{eq:rot_const}
    B_v = \frac{\hbar}{4 \pi \mu R_e^2}\quad.
\end{equation}
Using the Dunham model in \req{eq:E_dun}, we approximate the rotational constant as
\begin{equation}
B_v \approx 
Y_{01} + Y_{11} (v + 1/2)\quad,
\end{equation}
and the vibrational constant as
\begin{equation}
\omega_{v} \approx Y_{10} + Y_{20} (v + 1/2)\quad.
\end{equation}
With these approximations, we arrive at the equilibrium distances tabulated in \rtab{tab:literature_comparison} and the Franck-Condon factor estimates tabulated in \rtab{tab:fcf_all}.
These measurements confirm that the transition \XAtransitionvib{0}{0} and \XAtransitionvib{1}{1} with \remove{a} Franck-Condon factors of $f_{00} = 99.88\%$ and $f_{11} = 99.61\%$ are well suited for the proposed laser cooling schemes of AlCl \cite{Wan2016,Yang2016}. 
Applying the rigid rotor approximation to the potential energy surfaces from the {\it ab initio} theory, we find that our experimental results are in very good agreement with the theoretically predicted Franck-Condon factors..

\begin{table*}[ht]
    \caption{
    \label{tab:literature_comparison}Equilibrium constants for the $\Astate$ state in units of cm$^{-1}$ and bond lengths in units of \AA\ for AlCl followed from \rtab{tab:Us_fitted}. Also listed are the $\Xstate$ coefficients derived from \cite{Hedderich1993} for comparison.}
    \begin{ruledtabular}
    \begin{tabular}{llllllll}
        $T_e$ & $\omega_e$ & $\omega_ex_e$ & $B_e$ & $\alpha_e \times 10^3$ & $D_e \times 10^7$ & $R_e$(\AA) & Ref.\\\hline\\
        \multicolumn{8}{l}{$\Astate$}\\\hline
        38253.22(2) & 452.54(5) & 5.61(3) &  0.24535(2) & 2.652(7) & & 2.1220 & This work (\alcl{35}, exp.)\\
        38253.718 & 446.26 & 5.04 & 0.243132 & 3.0341 & 2.707 & 2.1326 & This work (\alcl{35}, theo.)\\

        38253.22(2) & 447.19(5) & 5.47(3) &  0.23958(2) & 2.559(7) & & 2.1220 & This work (\alcl{37}, exp.)\\
        38253.711 & 440.98 & 4.92 & 0.237412 & 2.928 & 2.582 & 2.1325 & This work (\alcl{37}, theo.)\\
        38267.55 & 441.6(2.3) & 2.81(37) & & & & & \cite{Mahieu1989}\\
        38254.0 & 449.96 & 4.37 & 0.259 & 0.006 &  & 2.06 & \cite{Herzberg2018}\\
        38436.3652 & 453.43 & 8.4793  & 0.2435 & & & 2.1324 & \cite{Xu2020} (theory)\\
        38224.44 & 455.60 &  & 0.24078 & & & 2.1330 & \cite{Wan2016} (theory)\\
        38303 & 471.81 & 9.61  & 0.2412 & & & 2.145 & \cite{Yang2016} (theory)\\\\
        \multicolumn{8}{l}{$\Xstate$}\\\hline
        0.0 & 481.77 & 2.10 & 0.243930 & 1.611 & 2.502 & 2.1301 & \cite{Hedderich1993} (\alcl{35})\\
        0.0 & 476.07 & 2.05 & 0.238191 & 1.555 & 2.385 & 2.1301 & \cite{Hedderich1993} (\alcl{37})\\
    \end{tabular}
    \end{ruledtabular}
\end{table*}

\section{Summary}

In conclusion, we present a direct comparison of high-level {\it ab initio} theory and high resolution absorption spectroscopy on the \XAtransitionvib{0}{0} and the \XAtransitionvib{1}{1} transitions of AlCl in a Helium buffer-gas cell. 
Our findings present an unprecedented level of quantitative understanding of AlCl, while improving the measured precision in the molecular bond length and other equilibrium constants of the $\Astate$ state by an order of magnitude, as listed in \rtab{tab:literature_comparison} along with a comparison of literature values.
Our theoretical model matches the measured transition frequencies very well without any adjustable parameters, except $T_e$, which is not uncommon in comparisons at this level of precision \cite{Puzzarini2013}.

From the experimental and theoretical results, we estimate the Franck-Condon factor of the \XAtransitionvib{0}{0} to be 99.88\%, as presented in \rtab{tab:FCs}, which confirms that AlCl is indeed an excellent candidate for laser cooling experiments.
In the future, we plan to study a beam of AlCl, produced from our buffer-gas source and apply radiative slowing to bring AlCl below the capture velocity of a molecular magneto-optical trap.

{\bf Acknowledgements.}
We acknowledge funding from the National Science Foundation (NSF) RAISE-TAQS program, grant number 1839153. AT and BKK acknowledge that part of this work was done under the auspices of the US Department of Energy under Project No. 20170221ER of the Laboratory Directed Research and Development Program at Los Alamos National Laboratory. Los Alamos National Laboratory is operated by Triad National Security, LLC, for the National Nuclear Security Administration of the U.S. Department of Energy (Contract No. 89233218CNA000001). \change{We would like to thank Daniel McCarron for helpful discussions.}

\appendix
\section{AlCl Line Centers}

In \rtab{tab:freqs35} and \rtab{tab:freqs37}, we list the line centers as measured, predicted with the Dunham fitting model, and calculated via {\it ab initio}.

\begin{table*}
    \centering
    \caption{ \label{tab:freqs37} \alcl{37} line center frequencies for the \XAtransition\ taken from a least-square fit of the observed lines to a Voigt function. Because the Q branch has multiple overlapping lines, we report the frequency of peak absorption for the Q branch observations. Theoretically predicted frequencies match the experiment after the $T_e$ adjustment (see text).}
    \begin{ruledtabular}
    \begin{tabular}{c c c c l c c c}
        $\mathbf{\nu}$ & $\mathbf{\nu}$' & J & J' & Expt. (cm$^{-1}$) & Dunham Model (cm$^{-1}$) & Theory adjusted (cm$^{-1}$) & Diff. (cm$^{-1}$) \\
         \hline
        0 & 0 & Q & * & 38237.6921(5) & 38237.69492 & 38237.68393 & -0.00819 \\
        \hline
        0 & 0 & 0 & 1 & 38238.1634(5) & 38238.16090 & 38238.15490 & -0.00852 \\
        & & 1 & 2 & 38238.6376(7) & 38238.63926 & 38238.62772 & -0.00993 \\
        & & 2 & 3 & 38239.1189(5) & 38239.11939 & 38239.10146 & -0.01742 \\
        & & 3 & 4 & 38239.6023(5) & 38239.60130 & 38239.57611 & -0.02620 \\
        & & 4 & 5 & 38240.0828(4) & 38240.08501 & 38240.05166 & -0.03115 \\
        & & 5 & 6 & 38240.5708(9) & 38240.57054 & 38240.52810 & -0.04268 \\
        \hline
        0 & 0 & 2 & 1 & 38236.737(1) & 38236.73642 & 38236.74200 & 0.00454 \\
        & & 3 & 2 & 38236.266(1) & 38236.26514 & 38236.27291 & 0.00714 \\
        & & 4 & 3 & 38235.796(1) & 38235.79567 & 38235.80477 & 0.00889 \\
        & & 5 & 4 & 38235.3283(5) & 38235.32801 & 38235.33758 & 0.00922 \\
        & & 6 & 5 & 38234.864(1) & 38234.86221 & 38234.87133 & 0.00743 \\
        & & 7 & 6 & 38234.4032(8) & 38234.39829 & 38234.40604 & 0.00283 \\
        \hline
        \hline
        1 & 1 & Q & * & 38201.9390(5) & 38201.93598 & 38201.91075 & -0.02827 \\
        \hline
        1 & 1 & 2 & 3 & 38203.3512(5) & 38203.35117 & 38203.30508 & -0.04608 \\
    \end{tabular}
   \end{ruledtabular}
\end{table*}

\begin{table*}
    \centering
    \caption{
    \label{tab:freqs35} \alcl{35} line center frequencies for the \XAtransition\ taken from a least-square fit of the observed lines to a Voigt function. Because the Q branch has multiple overlapping lines, we report the frequency of peak absorption for the Q branch observations. Theoretically predicted frequencies match the experiment after the $T_e$ adjustment (see text).}
    \begin{ruledtabular}
    \begin{tabular}{c c c c l c c c}
        $\mathbf{\nu}$ & $\mathbf{\nu}$' & J & J' & Expt. (cm$^{-1}$) & Dunham Model (cm$^{-1}$) & Theory adjusted (cm$^{-1}$) & Diff. (cm$^{-1}$) \\
         \hline
        0 & 0 & Q & * & 38237.4877(5) & 38237.49094 & 38237.48771 & 0.0 \\
         \hline
        0 & 0 & 0 & 1 & 38237.9691(5) & 38237.96818 & 38237.97001 & 0.00090 \\
        & & 1 & 2 & 38238.4589(3) & 38238.45802 & 38238.45418 & -0.00468 \\
        & & 2 & 3 & 38238.9509(4) & 38238.94967 & 38238.93926 & -0.01163 \\
        & & 3 & 4 & 38239.4426(3) & 38239.44312 & 38239.42526 & -0.01734 \\
        & & 4 & 5 & 38239.9388(2) & 38239.93841 & 38239.91216 & -0.02665 \\
        & & 5 & 6 & 38240.4364(5) & 38240.43555 & 38240.39996 & -0.03646 \\
        & & 6 & 7 & 38240.9341(3) & 38240.93458 & 38240.88864 & -0.04546 \\
        \hline
        0 & 0 & 2 & 1 & 38236.5086(4) & 38236.50944 & 38236.52313 & 0.01456 \\
        & & 3 & 2 & 38236.026(1) & 38236.02680 & 38236.04272 & 0.01695 \\
        & & 4 & 3 & 38235.5443(4) & 38235.54600 & 38235.56327 & 0.01896 \\
        & & 5 & 4 & 38235.0664(4) & 38235.06705 & 38235.08477 & 0.01836 \\
        & & 6 & 5 & 38234.5898(5) & 38234.58999 & 38234.60722 & 0.01745 \\
        & & 7 & 6  & 38234.115(1) & 38234.11484 & 38234.13064 & 0.01525 \\
        \hline
        \hline
        1 & 1 & Q & * & 38201.2184(5) & 38201.22056 & 38201.21839 & 0.0 \\
        \hline
        1 & 1 & 0 & 1 & 38201.705(2) & 38201.70510 & 38201.69754 & -0.00759 \\
        & & 1 & 2 & 38202.189(1) & 38202.18754 & 38202.17271 & -0.01592 \\
        & & 2 & 3 & 38202.6691(7) & 33202.66967 & 38202.64589 & -0.02324 \\
        & & 3 & 4 & 38203.1530(5) & 38203.15152 & 38203.11705 & -0.03594 \\
        & & 4 & 5 & 38203.630(1) & 38203.63310 & 38203.58620 & -0.04393 \\
        & & 5 & 6 & 38204.114(1) & 38204.11443 & 38204.05331 & -0.06095 \\
        \hline
        1 & 1 & 3 & 2 & 38199.7701(9) & 38199.77233 & 38199.77700 & 0.00687 \\
        & & 4 & 3 & 38199.2888(8) & 38199.28843 & 38199.29193 & 0.00314 \\
        & & 5 & 4 & 38198.8042(4) & 38198.80428 & 38198.80489 & 0.00070 \\
    \end{tabular}
    \end{ruledtabular}
\end{table*}


\bibliography{references,references-theory}

\end{document}